\documentclass[proof]{WileyASNA-v1}
\usepackage{epsfig}
\usepackage{upgreek}
\articletype{Original Paper}%

\received{<day> <Month>, <year>}
\revised{<day> <Month>, <year>}
\accepted{<day> <Month>, <year>}

\begin{document}

\title{Changing looks of the nucleus of the Seyfert galaxy NGC~1566 compared with other changing-look AGNs
\protect\thanks{Changing looks of the nucleus of the Seyfert galaxy NGC~1566}
}

\author[1]{Victor Oknyansky*}

\authormark{OKNYANSKY}

\address{\orgdiv{Sternberg Astronomical Institute}, \orgname{M. V. Lomonosov Moscow State University}, \orgaddress{Universitetsky pr-t, 13, Moscow,  119234}, \country{Russia}}

\corres{*Corresponding author: V.~L.~Oknyansky
\email{oknyan@mail.ru}}


\abstract[Abstract]{We  present results of a long-term optical, UV and X-ray study of variability of the nearby changing-look (CL) Seyfert NGC~1566 which was observed with the {\it Swift} Observatory  from 2007 to 2020. We summarize our previously published  spectroscopic and photometric results and present new observations.   We reported on the alteration in the spectral type of NGC 1566 in 2018 (REF1). Moreover, we focused on the exceptional postmaximum behavior after 2018 July, when all bands dropped with some fluctuations (REF2).  We observed four significant re-brightenings in the post-maximum period.  We have found differences in X-ray and UV/Optical variability.   The $L_{\rm uv}/L_{\rm x}$ ratio was decreased during 2018-2020.  New post-maximum spectra covering the period 2018 November 31  to 2019  September 23 show dramatic changes compared to 2018  August 2, with fading of the broad lines and [Fe X] $\lambda$6374 until 2019 March  (REF2).   Effectively, two changing look (CL) states were observed for this object: changing to type 1.2 and then returning to the low state  as a type 1.8~Sy. We suggest that the changes are mostly due  to fluctuations in the energy generation. 

Variability properties of NGC1566 are compared with our results for other CL active galactic nuclei (AGNs).}


\jnlcitation{\cname{%
\author{Oknyansky, V.~L.}, 
 } (\cyear{2021}), 
\ctitle{Changing looks of the nucleus of the Seyfert galaxy NGC~1566}, \cjournal{Astron. Nachr.}, \cvol{ \url{https://doi.org/10.1002/asna.20210080}}.}


\maketitle

\section{Introduction}\label{sec1}

The NGC1566 was discovered in 1826 by James Dunlop but intensive investigations of it  were triggered after discovery of the Seyfert phenomenon in this object in 1956 \citep{Vaucouleurs1961,Vaucouleurs1973}  which was  confirmed in 1962 \citep{Shobbrook1966}. 
These investigations of NGC~1566 have long history (see e.g. references at  \cite{Silva2017, Oknyansky2019, Oknyansky2020c, Parker2019}). 
The NGC1566 is one of the brightest  and nearest galaxies with AGN in the South Hemisphere. This object is also nearest  Changing Look (CL) AGN \citep{Oknyansky2018b,Oknyansky2019}. It is one of two  (also NGC3516)  firstly discovered (soon after discovery of variability of AGNs in continuum by \cite{Fitch1967}) CL AGNs \citep{as68,Pastoriza1970}, however  this designation came into common use only in the past decade. The CL AGNs are active nuclei of galaxies which undergo dramatic variability of the emission line profiles and classification type, which can change from type 1 (showing both broad and narrow lines) to type 1.9 (where the broad lines almost disappear) or vice versa within a short time interval (typically a few months). This dramatic spectral change in the NGC1566 was correlated with optical variability \citep{Q1975}. Due to the optical monitoring since 1954  about 20 years (first published optical light curve) the object was in bright state during few years till 1962 then it shortly dropped down and shortly moved up to the maximal level at 1963. Such re-brightenings after strong outbursts and CLs probably are typical property of CL AGNs and were fixed in several other ones of them \citep{Oknyansky2017a, Oknyansky2018b, Katebi2019}. After 1963 the NGC~1566  was  mostly  in low state  what was a reason to note it  as a ``weak Seyfert", however occasional recurrent brightenings of H$\upbeta$ and optical continuum (not as strong as the outbursts of 1962 and 2018) were observed in the period 1982-1991 \citep{Alloin1985, Alloin1986, Kriss1991, Winkler1992, Baribaud1992}.

The first multi-wavelength (X-ray, UV, optical and IR), UV/optical emission lines investigations of the variability of NGC~1566, as well as the first IR reverberation mapping, were published by \cite{Baribaud1992}. It is seems that found IR time delays ($\sim$ several months) was significantly overestimated (see discussion of that in Paper II) and more realistic value for the IR time delay is less than 20 days \citep{Oknyansky2001}.
 
NGC~1566 is a galaxy with a very well-studied variable active nucleus, however the spectral and photometric data have big gaps. The most intensive multiwave photometic observations were done during past years after discovery of a new reawakening \citep{Kuin2018, Ferrigno2018, Grupe2018, Parker2019, Cutri2018} and new CL phase (see \cite{Oknyansky2018b, Oknyansky2019, Oknyansky2020c} and references there). During these few years 2 CLs were discovered in the object: first happened in 2018 when object changed from Sy1.9 to Sy1.2 (\cite{Oknyansky2019} = Paper I) and second  one in 2019 when object changed back to the low state (\cite{Oknyansky2020c} = Paper II). This was happened about 60 years after previous  CLs  \citep{Pastoriza1970, Q1975}, but, taking into account too long gaps in the spectral coverage,  some other possible CLs could well have been missed. Meanwhile it is entirely possible that  CLs of NGC~1566 such as in 1962  and 2018 (see \citealt{Oknyansky2018b} and Paper I) are recurrent events happening on time-scales of several decades. 

The spectral transition during the CL events (during 2018-2020) of NGC~1566 not only manifested in dramatic intensity changes of the broad emission lines but also in significant strengthening of the  UV Balmer continuum and high-ionisation coronal lines such as [FeX] $\lambda$6374 (Papers I -- II). Such variability of coronal lines was observed also in some other CL AGNs: NGC~4151 \citep{Oknyansky1982, Chuvaev1989,Oknyansky1991} and NGC~3516 (in preparation). The  H$\upalpha$/H$\upbeta$ ratio (for broad components) changed from $\sim$2.6 in high state  to $\sim$6.2 in the low state (see details in Paper II). The variability of the Balmer decrements is a typical property of a CL AGN  \citep[see e.g.,][]{Shapovalova2004, Shapovalova2008,Gaskell17,Ilich2012,Jaffarian2020}. 
After maximum was reached in July 2018, fluxes in all bands declined, with some  re-brightenings Event 1 in December 2018 \citep{Grupe2018b}, Event 2 in the end of May 2019 \citep{Grupe2019}, Event 3 in August 2019 (Paper II) and Event 4 in May 2020 \citep{Jana2021}. This decline was accompanied by a decrease of the $L_{\rm uv}/L_{\rm x}$ ratio (Paper II).  The estimated Eddington ratios were about 0.055$\%$ for minimum in 2014 and  2.8$\% $ for maximum in 2018.
  
In this paper we present results of the continuation of our multi-wavelength (optical, UV and X-ray) monitoring of NGC~1566 using the data obtained with {\it Neil Gehrels Swift Observatory} over the period 2007-2020 combined them with published before results in PaperI and Paper II, which include also spectral data obtained with the South African Astronomical Observatory 1.9-m telescope from 2018 August  to 2019 September and photometry with the MASTER Global Robotic Network from 2014 to 2020 December. We compare these results with the revealed before facts for other CL AGNs to fix some common properties which can eventually help us to understand better physics of the CL phenomenon.  
  
\section{Observational data and results}
\begin{figure}[h!]
\centering 
\includegraphics[width=0.48\textwidth,angle=0]{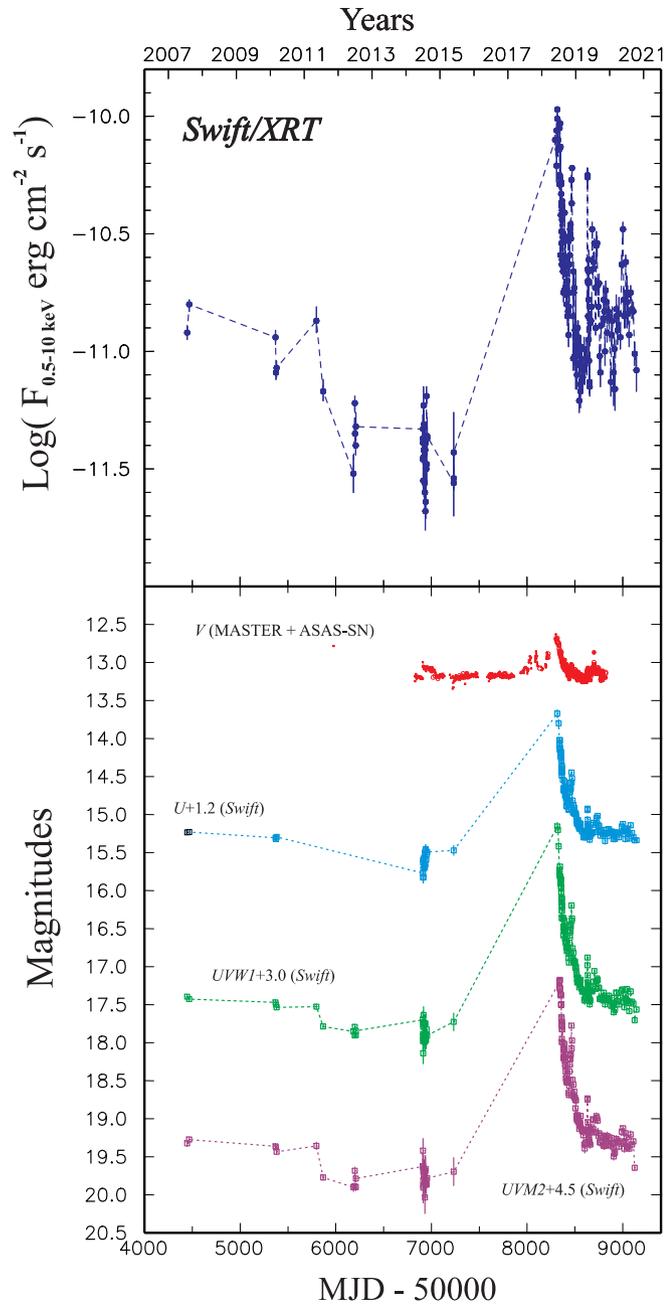}
\caption {Multi-wavelength observations of NGC~1566 spanning the period 2007 December 11 to 2020 October 21. {\it Top panel:} The {\it Swift}/XRT 0.5--10 keV  X-ray flux (in erg cm$^{-2}$ s$^{-1}$). {\it Bottom panel:} Optical--UV photometric observations. The large open circles represent MASTER unfiltered optical photometry reduced to the $V$ system while the points are $V$ ASAS-SN (nightly means) reduced to the {\it Swift} $V$ system. The filled circles show MASTER $V$-band photometry. The small open boxes correspond to the {\it U}, {\it UVW1} and {\it UVM2} {\it Swift}/UVOT photometry.\label{fig1}}
\end{figure}

\begin{figure}[h!]
\centering 
\includegraphics[width=0.48\textwidth,angle=0]{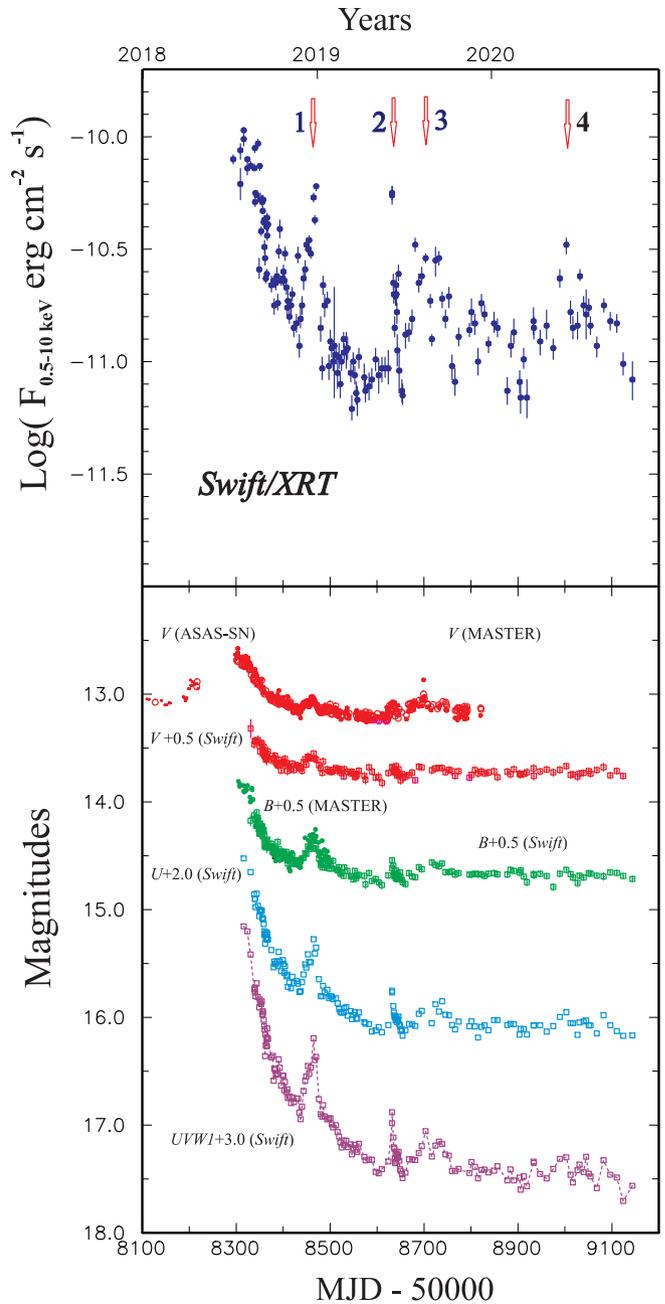}
\caption{Multi-wavelength observations of NGC~1566 shown just for 2018-2020. {\it Top panel:} The {\it Swift}/XRT 0.5--10 keV X-ray flux (in erg cm$^{-2}$ s$^{-1}$). {\it Bottom panel:} The large open circles represent MASTER unfiltered optical photometry reduced to the $V$ system while the points are $V$-band ASAS-SN (nightly means) reduced to the {\it Swift} $V$ system. The filled circles show MASTER $BV$ photometry results. The open boxes correspond to the {\it UVW1} and {\it UBV} data obtained by {\it Swift}. The arrows indicate the Events 1, 2, 3  and 4 (see text for the details).\label{fig2}}
\end{figure}

NGC~1566 has been regularly observed by the {\it Neil Gehrels Swift Observatory} \citep{Gehrels2004} for many years, starting in late 2007  and the results of the monitoring  were published mostly  by  \cite{Kawamuro2013, Ferrigno2018, Grupe2018, Grupe2018b, Oknyansky2019,  Oknyansky2020c, Grupe2019, Parker2019, Jana2021}. In the present paper we have added to the analysis the most recent data (both from the XRT and UVOT telescopes), using the same methods as in Paper I and II, but we uniformly re-reduced all available data to ensure usage of the most recent versions of the software and calibration files.  No significant variations were found between different versions of the reduced X-ray data. Meanwhile last update of the UVOT calibrations can give the difference up to almost 0.3 mag  (see details here ~ ~\url{http://www.swift.ac.uk/analysis/uvot/}). The new data include 28 dates for between 5th December 2019 and 20th October 2020. The data allows us to trace better the long and shot term  evolution of NGC~1566's behaviour including  post maximum period. 

We summarize a study of optical, UV and X-ray light curves of the nearby changing look active galactic nucleus in the galaxy NGC 1566 obtained with the Neil Gehrels Swift Observatory and the MASTER Global Robotic Network over the period 2007-2019 published before in Paper II and new data for the years 2019-2020 (see Figures~\ref{fig1}--~\ref{fig2}).  The MASTER  observations  and  results of the ASAS-SN \citep[All-Sky Automated Survey for Supernovae, ][]{Shappee2014, Kochanek2017, Dai2018} $V$-band magnitudes reduced to the $V$ {\it Swift}/UVOT system are presented in Figures~\ref{fig1}--~\ref{fig2} (till 2019 December 5) as were published before in Paper II. The light curves in the optical {\it UBV}, {\it UVW1}, {\it UVM2} and X-ray bands are well correlated as can be clearly seen in Figure~\ref{fig1} (bottom panel) for 2007-2020 and Figure~\ref{fig2} (bottom panel) for 2018--2020.  The light curves till
2019 December 5  were discussed in details in the Paper II. There were considered 3 re-rebrightening Events 1-3. Here in new data we see one more Event 4 (2020 August), which was firstly noted by \cite{Jana2021}. The Event 3 was most significant and it happened about one year after the main maxim but was seen mostly from the MASTER data since its maximum  was missed by {\it Swift} observations. We didn't see any clear signature of periodicity of these events as it was suspected by \cite{Jana2021}. The event 2 and 3 were very fast and short and so some other such re-brightening could well have been missed.  In the Paper II was firstly found a decrease of the $L_{\rm uv}/L_{\rm x}$ ratio at post-maxim decrease. Here that is seen more clear  from the Figures~\ref{fig1}--~\ref{fig2}:  the X-ray flux was in mean grow up after the minimum in  2019 March but UV brightness had tendency  to move down in mean. This effect for UV bands is significant, it is smaller for $U$ and not seen for $BV$ variations.

\section{Discussion}

The properties of multi-wavelength variability of  NGC 1566 is not unusual for a CL AGNs. The dramatic changes of factors  few tens  in the X-ray flux (see Paper I) along with correlated variations at UV and optical wavelengths is a typical property of CL AGNs. However, such a strong correlation of the X-ray and UV/Opt is not common among  AGNs \citep[see e.g.,][] {Edelson2000, Buisson2017},  there are exceptions for some other CL AGNs, viz. NGC~ 2617, NGC~4151, NGC~3516, and NGC~5548 \citep[see e.g.,][]{Shappee2014, Oknyansky2017a, Edelson2017, Mchardy2014, Oknyansky2021}. The UV/Optical variations are typically  delayed relative to the X-ray fluctuations by about a few days. These  delays  correlate with the masses of the supermassive black holes (SMBH). From these results we can conclude that the variability across several wavebands in CL AGNs (spanning from X-rays to the  UV/Optical) is driven by variable illumination of the accretion disc (AD) by soft X-rays. We have to determine the most significant common features of such variability.  That can help us to find the most likely explanations for the CL phenomenon (see e.g. discussion and references in Papers I--II and below).  There are several different possibilities that have been considered: variable obscuration, AD flares, tidal disruption events (TDEs), and supernova event  (see e.g. more detailed discussion and references in Papers I -- II  and below) .
In case of NGC~1566 the duration of the rise and decay  times  for the  main brightening in 2018 was about the same ($\sim 9$ months) and much longer than the recurrent brightenings (with lower amplitude) described by \cite{Alloin1986} ($\sim$20 days).  A similar dramatic brightening was observed previously in 1962, with re-brightening about one year later \cite{Quintana1975}. Such re-brightenings soon after the main maximum may be common in the variability of CL AGNs.   
Whilst there is good evidence for obscuration in many of AGNs (see \citealt{Gaskell17, Jaffarian2020} and references therein), it is not the dominant variability mechanism.  It is difficult to explain the dramatic rise of the X-ray and UV/optical flux by some variable obscuration. The main problem with the obscuration theory is that it would predict much longer time scales for the spectral changes than what is observed in CL AGNs \citep[see e.g.,][]{Sheng2017}. For NGC 1566 we expect timescales of more than a few years for a CL event in the obscuration scenario, which is much longer than what is observed.  It does not exclude the possibility of intervening clouds causing variable obscuration, but seems to exclude that as a dominant explanation for the CL mechanism. The more obvious explanation of this, and AGN variability in general, is that it is due to variation of the intrinsic energy-generation rate.
These two leading possibilities -- intrinsic variability and variable obscuration -- do not need to be mutually exclusive. For example, a dramatic change in the energy generation could cause the sublimation of dust in clouds near the central source energy source \citep{Oknyansky2017a, Oknyansky2019a}. If such sublimation occurs in some clouds along the line of sight, then the rise in the UV can be explained in part with the change in obscuration. The typical time for the recovery of dust clouds after the UV flux has abated can be several years \citep{Oknyansky2017a, Oknyansky2019a,  Kishimoto2013}, but it could be less for NGC~1566 if we take into account the significantly smaller mass of the SMBH  compared with NGC~4151. So we could see some increase in obscuration during the time when the energy generation is falling.   
We do see some differences in variability in different wavelengths after Mar. 2019, which can be explained by variable obscuration since it must be most significant for the UV and less in optical bands, and with hardly any noticeable effect for X-ray flux. Similar decrease  in UV/X-ray ratio at post maximum deeming was found by us for NGC~2617 and NGC~3516 (in preparation). That is in agreement with  absorbing column variations (more than of a factor of 3 from during 2015 -- 2018) found in the NGC~1566 by \cite{Jana2021} (however the estimates are model-dependent and the variations can be due to dust sublimation).  After the maximum in 2018 the estimated (by Jana et al.) hydrogen column density had tendency  to grow up in correlation with variations of H${\alpha}$/H${\beta}$ ratio (found in Paper II).   Meanwhile there are  some other possible explanations for these variations and difference in X-ray and UV variability. For example, the strong change in the $L_{\rm uv}/L_{\rm x}$ flux ratio observed  in the NGC~1566 after Mar. 2019 (see Paper II) can be connected with changing of the height of the X-ray source that can reduce the amount of UV/optical emission produced by reprocessing in the AD \citep[see e.g.,][]{Breedt2009}. Also a transition in accretion mode might explain the observed differences in properties of  variability in  high and low states \citep[see e.g.,][]{Zhu2020}
The most intriguing question is what the mechanism would be for such dramatic switching on or off for the energy generation of CL AGNs. This has been explored in some firsts reports on CL events as well as in recent papers \citep{Lyuty1984, Penston1984, Runnoe2016, Katebi2019}. See discussion of the possible mechanisms for CLs in  \cite{ Lightman1974, Parker2019, Sniegowska2021, Zhu2020, MacLeod2019,Runnoe2016, Ruan2019} and Paper II.  The TDE and supernova explanations can be rejected given the differences between expected and observed time scales for such events, as well as their spectral properties and evolution.  An alternative theory proposes the tidal stripping of stars \citep{Campana15, Ivanov-Chernyakova06}, which could lead to more frequent and recurrent events \citep{Komossa17} if these stars have bound orbits similar to some known objects near the SMBH in the Milky Way.  However, the mechanism has not been investigated in sufficient details yet. At present we are far from understanding the CL phenomenon, and many questions remain still.

\section{Conclusions}
We summarize a study of the NGC~1566 published before in Papers I--II with new optical, UV and X-ray light curves  with the Neil Gehrels Swift Observatory over the period 2007-2020 taking into account most recent versions of the software and calibration files.
We found some common properties  in variability of the NGC~1566 with  other CL AGNs which can be useful  for the future  models intended to explain the CL phenomenon: 

1. CLs are recurrent events.

2. After the strong outbursts, as a rule some re-brightenings are happened.

3. The $L_{\rm uv}/L_{\rm x}$ flux ratio is significantly decreased at post maximum time.

4. The UV/Optical variations are delayed to X-ray ones  on few days.

5 The UV and X-ray variations are correlated but some uncorrelated events or trends are typical.

6. During the strong outbursts  dust sublimates in the clouds. The dust can be recovered if the object moves to a low state at least on few years. That can explain variations of obscuration.

7. The H$\alpha$/H$\beta$ is higher in low states (see e.g. \citealt{Shapovalova2010} and Paper II) .

8. Variability of  coronal lines might be common property for CL AGNs.

9. CL AGNs have low accretion rate << 1\% Edd in minimum and few 
\% in maximum. 

10. The significant strengthening of the  UV Balmer continuum might be common for CL AGNs since were
noted in  NGC~2617 \citep{Shappee2014}, NGC~3516 \citep{Oknyansky2021} and some other CL AGNs.






\section*{Acknowledgments}
 I  express my thanks to the {\it Swift} ToO team for organizing and executing the observations and to S.S.~Tsygankov for help with the {\it Swift} data processing. I am thankful to my collaborators in the Paper II and in \cite{Oknyansky2021} for their fruitful work which made this publication possible. This work was supported in part by M.V.Lomonosov Moscow State University Program of Development.  I am grateful to H.~Netzer,  C.M.~Gaskell, J.M.~Wang, A.M.~Cherepashchuk, A.V.~Zasov, I.F.~Bikmaev  and P.B.~Ivanov   for useful discussions. I am thankful to J.~Jadeja for editing of the text.

\bibliography{1566}%
\section*{Author Biography}

\begin{biography}
{\includegraphics[width=60pt,height=70pt]{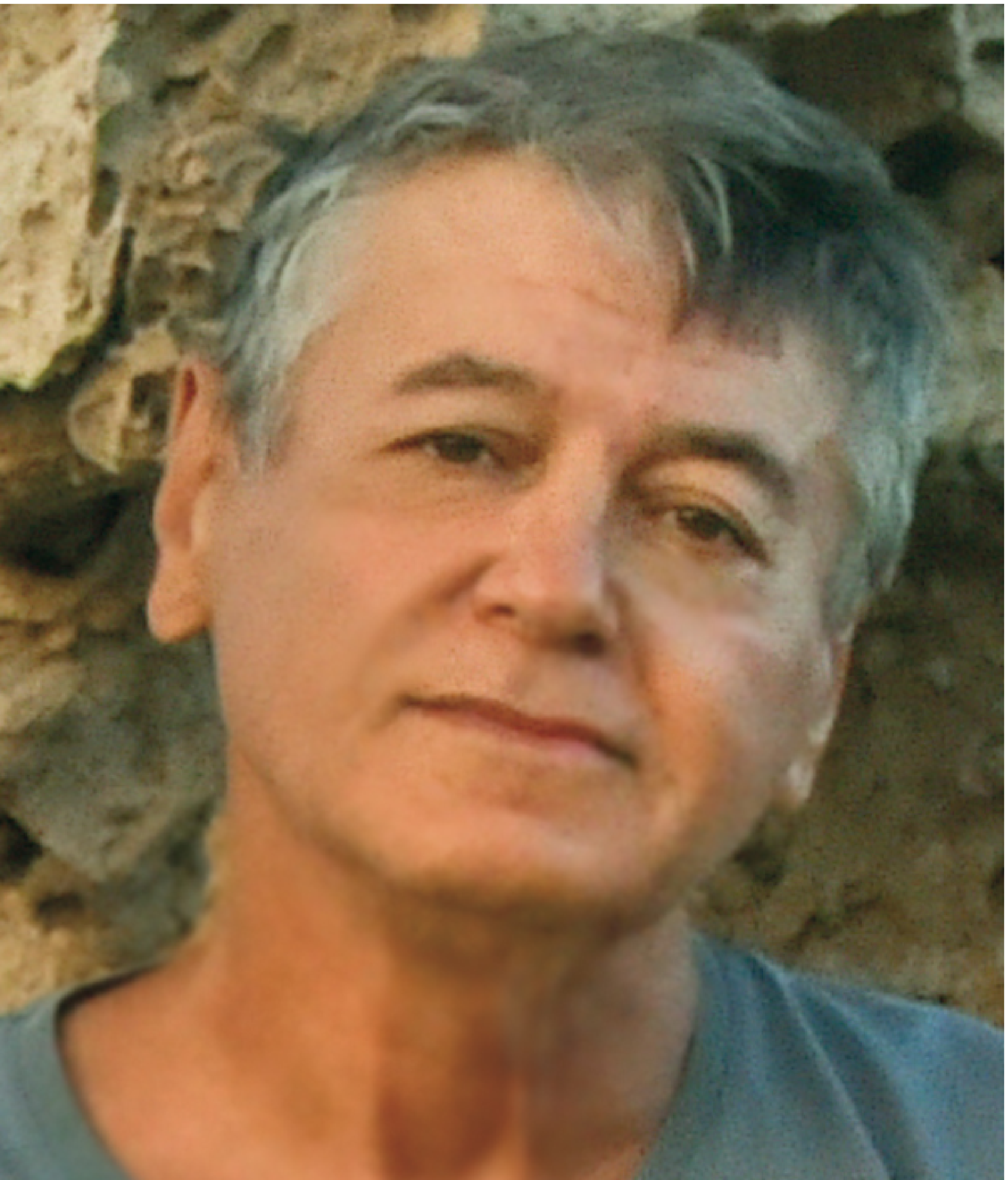}}{\textbf{Victor Oknyansky.} 
Victor Oknyansky graduated from Odessa State University in 1976  (a bachelor of astronomy).  While he was still a student, he began to study photometric and spectral  variability of active galactic nuclei, he completed his diploma practice at the Crimean Astrophysical Observatory, where he began to work closely with  Prof. Victor Lyuty and  Prof. Konstantin Chuvaev. He was published many scientific works in collaborations with them. V. Lyuty was a supervisor in the research on the topic of his thesis “Study of cyclic phenomena in the variability of active galaxies and quasars”, which he defended in 1986 (SAI MSU) and got the Ph.D. degree. 1998-1990 he worked as a senior researcher in the Special Astrophysical Observatory as a visitor astronomer. After moving to Moscow in 1991,   he started to work  at the SAI MSU,   where  he currently works as a senior research associate. He received a grant from the Swedish Institute for Research at the Uppsala Astrophysical Observatory during 1994-1995 (host Prof. Ernst van Groningen), as well as a grant from the British Royal Society for long-term studies (1999–2000) together with British scientists at the University of San Andrews (Prof. Keith Horne.    His main research interests are related to studies of the variability of AGNs, quasars, CL AGNs, and gravitational lenses. He was published tens of articles  in leading scientific journals, including journals with a high impact factor: Ap.J., MNRAS, AsAp., Ap.SpSci. Repeatedly he was giving oral and invited talks at international conferences, including such countries as the USA, Australia, Israel, Japan, France, Spain, Belgium, Italy, China, Ethiopia. Past 5 years he was mostly studied variability of the CL AGNs in collaboration with Professors  Michael Brotherton, Sergey Tsygankov, Jian-Min Wang, Vladimir Lipunov, Hartmut Winkler, Dragana Ilic, Martin Gaskell as well as with  collaborators from the SAI MSU.

}
\end{biography}

\end{document}